\documentclass{iopart}

\usepackage{graphicx,iopams}

\newcommand{\p}[3][]{\frac{\partial^{#1}#2}{\partial{#3}^{#1}}}
\newcommand{\Pe}{\mbox{P\hspace{-1pt}e\hspace{1pt}}}
\newcommand{\dt}{\Delta t}
\newcommand{\dx}{\Delta x}
\newcommand{\dz}{\Delta z}
\newcommand{\dxsq}{\langle\Delta x^2\rangle}
\newcommand{\dzsq}{\langle\Delta z^2\rangle}

\newcommand{\tuzun}{T\"uz\"un\ }

\begin{document}

\title[Velocity profile of granular flows inside silos and hoppers]
{Velocity profile of granular flows inside silos and hoppers}

\author{Jaehyuk Choi,$^1$ Arshad Kudrolli,$^2$ and Martin Z. Bazant$^1$}  
\address{$^1$ Department of Mathematics, Massachusetts Institute of Technology, Cambridge, MA 01239, USA}
\address{$^2$ Department of Physics, Clark University, Worcester, MA 01610, USA}
\ead{jaehyuk@math.mit.edu,akudrolli@clarku.edu,bazant@math.mit.edu}

\begin{abstract}
We measure the flow of granular materials inside a quasi-two
dimensional silo as it drains and compare the data with some existing
models. The particles inside the silo are imaged and tracked with
unprecedented resolution in both space and time to obtain their
velocity and diffusion properties. The data obtained by varying the
orifice width and the hopper angle allows us to thoroughly test models
of gravity driven flows inside these geometries. All of our measured
velocity profiles are smooth and free of the shock-like
discontinuities (``rupture zones'') predicted by critical state soil
mechanics. On the other hand, we find that the simple Kinematic Model
accurately captures the mean velocity profile near the orifice,
although it fails to describe the rapid transition to plug flow far
away from the orifice. The measured diffusion length $b$, the only
free parameter in the model, is not constant as usually assumed, but
increases with both the height above the orifice and the angle of the
hopper. We discuss improvements to the model to account for the
differences. From our data, we also directly measure the diffusion of
the particles and find it to be significantly less than predicted by
the Void Model, which provides the classical microscopic derivation of
the Kinematic Model in terms of diffusing voids in the
packing. However, the experimental data is consistent with the
recently proposed Spot Model, based on a simple mechanism for
cooperative diffusion. Finally, we discuss the flow rate as a function
of the orifice width and hopper angles. We find that the flow rate
scales with the orifice size to the power of $1.5$, consistent with
dimensional analysis. Interestingly, the flow rate increases when the
funnel angle is increased.
\end{abstract}

\pacs{45.70.Mg, 66.30.-h}

\submitto{\JPCM special issue on Granular Media}

\maketitle

\section{Introduction}
\label{sec:intro}
Granular materials display a surprisingly complex range of properties
which make them appear solid or liquid like depending on the applied
conditions~\cite{duran00,jaeger96}. Because the interaction between
the grains is dissipative and the thermal energy scale is small
compared with the energy required to move grains, such materials
quickly come to rest unless external energy is supplied
constantly. Although vibro-fluidization and tumbling~\cite{ottino02}
is frequently used to excite granular materials, flows driven purely
by gravity can occur in nature as well.  Typical granular flows are
dense and a fundamental statistical theory is not avaliable to
describe their properties. One reason for this situation is the lack
of quantitative data which can be used to test and develop models of
dense granular flow. In this paper, we focus on flows inside silos and
hoppers in order to elucidate the nature of the flow and to test
existing models. Such systems are ubiquitous due to the need to store
and process granular materials in devices ranging from simple hour
glasses to sophisticated nuclear pebble
reactors~\cite{talbot02,kadak04}.

Several aspects of granular drainage have been studied over the
years. Beverloo thoroughly investigated the relation between the
orifice size and the mass flow rate in cylindrical silos and proposed
a formula describing the observed dependence~\cite{beverloo61}. 
Using radiography, Baxter, \etal observed the density wave in the hopper
flow and showed various patterns of the wave depending on the particle 
roughness and the hopper angle~\cite{baxter89}.

The velocity field of the flow inside a silo has been described by two
different approaches. One is based on the critical-state theory of
soil mechanics which relates stress and density to predict velocity
field or mass flow rate~\cite{prakash91, nedderman92}. Although this
approach has the appeal of starting from mechanical considerations,
some questionable assumptions are made to resolve indeterminacy in the
stress tensor, and the resulting equations are mathematically ill-posed
and can lead to violent singularities~\cite{schaeffer87,pitman87}. The
solutions available for hoppers possess shock-like velocity
discontinuities (``rupture zones'')~\cite{nedderman92}, which are not
seen in our experiments (see below).

The second approach ignores the stress field and attempts a purely
kinematic description of the velocity profile, starting from an
empirical constitutive law.  A theory of this type was first discussed by
Litwiniszyn, who introduced a stochastic model in which particles
perform random walks through available ``cages''~\cite{lit58, lit63a,
lit63b}. Later, Mullins independently proposed an equivalent
stochastic model of the flow in terms of ``voids" and extensively
developed the continuum limit, where a diffusion equation
arises~\cite{mullins72, mullins79}. Decades later, Caram and Hong
revisited the Void Model and implemented it explicitly in computer
simulations on a triangular lattice (where the voids are simply
crystal vacancies)~\cite{caram91}.  

As an alternative to the microscopic void picture, Nedderman and
\tuzun derived the same continuum equation starting from a
constitutive law relating horizontal velocity and downward velocity
gradient~\cite{nedderman79, tuzun79}. Regardless of its derivation,
the Kinematic Model predicts velocity fields with only one free
parameter. In light of its simplicity, early experiments on silo
drainage were viewed as successes of the
model~\cite{tuzun79,mullins74,tuzun82b}, even though it has since
fallen from favor in engineering~\cite{nedderman92}. Although the free
parameter has been observed to be proportional to grain diameter in
all experiments, the constant of proportionality does not
agree~\cite{tuzun79,mullins74,samadani99}. Furthermore, Medina,
\etal~\cite{medina98b} have reported that the kinematic parameter 
varies within a silo when the flow is analyzed in detail by particle image
velocimetry.

In addition to the studies of the flow pattern, the diffusion of
particles has been investigated as well. Hsiau and Hunt~\cite{hsiau93}
and Natarajan, \etal~\cite{natarajan95} imaged tracer particles in a
dense flow inside a vertical channel with various boundary wall
condition to investigate the concept of ``granular temperature''. From
an analysis of velocity fluctuations, they found that particles shows
normal diffusion and that the diffusivity in the stream wise direction
is higher than in the transverse direction. Later, Menon and Durian
used diffusing-wave spectroscopy to measure the dynamics of
$100\mu{\rm m}$ glass beads inside a three-dimensional flow with
improved temporal resolution, albeit at rather small length
scales~\cite{menon97}. They reported that the particles show ballistic
flight between collisions over a short time scale, and normal
diffusion over longer time scale, although the collision distance of
28 nm (1/10,000 of a grain diameter) could perhaps be associated with
sliding or rotating asperities in frictional contacts. In any case,
the randomizing gas-like collisions assumed in kinetic
theories~\cite{savage79,jenkins83,hsiau93b} have not been confirmed in
any experiments on dense flows.

With rapid advances in high-speed digital imaging technology, it is
now possible to simultaneously record thousands of individual
particle positions with high spatial and temporal resolution. In a
recent experiment by our group using this technique, the dynamics of 3
mm glass beads near a transparent wall in a three-dimensional silo was
observed to be sub-ballistic but super-diffusive over short time
intervals, and diffusive over long time intervals~\cite{choi04}. The
data was argued to be consistent with slow cage rearrangement with
particles remaining in long-lasting contacts by showing that the
diffusion scaled only with distance traveled.

Therefore inconsistencies can be noted in reported results which need
to be resolved with thorough investigations. In the next section, we
introduce the kinematic description of granular flow in silos and
hoppers in detail. Since the stress field is not measured by imaging
techniques, we do not assess critical-state mechanical models, aside
from seeking the presence of the predicted shocks in the velocity
field.  Then, we outline the experimental setup in
\sref{sec:exp}, and compare the prediction of models with our
experiments in \sref{sec:analysis}. We discuss the implications of the
comparison in \sref{sec:discussion} and finally summarize the results
in \sref{sec:summary}.

\section{Models for the mean velocity profiles}
\label{sec:model}

A simple kinematic description of the mean velocity profile in silos
and hoppers has been developed by since the 1950s, from a variety of
theoretical perspectives~\cite{nedderman92}. The continuum Kinematic
Model starts from an empirical constitutive law relating velocity
components~\cite{nedderman79}, which can be derived as a continuum
limit of the (earlier) Void Model~\cite{lit58,mullins72}. The latter
is a more complete theory, because it provides a microscopic mechanism
for flow, which can be checked by experiments on diffusion and
mixing. Recent experiments, however, have firmly rejected the void
hypothesis. On the other hand, an alternative stochastic description,
the Spot Model~\cite{bazant04a,bazant04b}, which starts from a
cooperative mechanism for random-packing rearrangements, roughly
preserves the mean flow profile of the Kinematic Model, with much less
diffusion and slow cage breaking, consistent with
experiments~\cite{choi04}.

\subsection{The Kinematic Model}
\label{sub:kinematic}

Nedderman and \tuzun~\cite{nedderman79, tuzun79} proposed a model
based on the following constitutive law relating velocity components:
\begin{equation}
  \label{eq:consti}
  u = b \p{v}{x},
\end{equation}
which states that the horizontal velocity $u$, is proportional to the
horizontal gradient (i.e. the shear rate) of the downward velocity
$v$. This assumption is based on the fact that particles tend to drift
horizontally towards a region of faster downward flow as they are
likely to find more space to move in that direction. Assuming that the
density fluctuation is small in dense granular regimes, they combined
Eq.~\eref{eq:consti} with the incompressibility condition,
\begin{equation}
  \label{eq:incompress}
  \p{u}{x} - \p{v}{z} = 0,
\end{equation}
and obtained an equation for the downward velocity,
\begin{equation}
  \label{eq:pde_v}
  \p{v}{z} = b\p[2]{v}{x}.
\end{equation}
Eq.~\eref{eq:pde_v} has a form of the diffusion equation, where time is
replaced by the vertical coordinate $z$. When an ``initial condition"
is given for $v$ at the bottom of the silo at $z = 0$, the velocity
diffuses upward. The boundary condition assumed at the side walls of
the silo is that the velocity is parallel to the wall. Although the
authors did not discuss this situation, the condition can be naturally
generalized to the case where the side walls are not vertical. It is
written as 
\begin{equation}
  \label{eq:bdd} u\, n_x - v\, n_z = 0 \quad {\rm at}\quad (x,z) {\rm
  \;\;on\;\; the\;\;side\;\;wall},
\end{equation}
where $(n_x, n_z)$ is the normal vector at the boundary.

For a semi-infinite quasi-two dimensional system ($-\infty < x <
\infty$) with a point-like orifice at $z=0$ which acts as a source of
velocity, a similarity solution exists:
\begin{equation}
  \label{eq:similarity}
  v(x,z) = \frac{Q}{\sqrt{4\pi b z}}\; e^{-x^2/4bz},
\end{equation}
where, $Q$ is the flow rate per unit thickness of the silo. We refer
to the constant of proportionality $b$ in Eq.~\eref{eq:consti}, as the
``diffusion length," as it is has units of length.
We provide microscopic understanding of $b$ in \sref{sub:void}.

The Kinematic Model has been tested experimentally, and the parameter,
$b$ has been measured by various groups.  Nedderman and \tuzun
observed $b \approx 2.24d$ for various particle size~\cite{tuzun79}.
Experiments by Mullins with monodisperse iron ore particles imply $b
\approx 2d$~\cite{mullins74}.  Medina, \etal used the particle image
velocimetry (PIV) technique to obtain the velocity field and found
that the diffusion length increases from $b \approx 1.5d$ to $b \approx 4d$
as the height increases to fit the field.~\cite{medina98b}. Samadani,
\etal reported $b \approx 3.5d$ for monodisperse glass beads using
difference imaging to find velocity contours~\cite{samadani99}.  All
the groups claimed that the prediction of the Kinematic Model
qualitatively agree with their experiment. The fact that a single
fitting parameter $b$ suffices to reproduce the entire flow field
should be viewed as a major success of the Kinematic Model.

In order to test the Kinematic Model more thoroughly, we use numerical
methods to solve the Kinematic Model subject to the same dimensions
used in our experiments. For this purpose, we define the stream
function, $\psi(x,z) = \int_{0}^x v(s,z) ds$ and solve for $\psi(x,v)$
rather than $v(x,z)$. Formulated in terms of $\psi$, the boundary
condition turns into a Dirichlet one from a rather complicated one
given by Eq.~$\eref{eq:bdd}$. Furthermore, it is more convenient for
the hopper geometry with inclined boundaries. If the width of the
system is given by $L(z)$ and the silo is symmetric about its center
[e.g. $-L(z)/2\le x \le L(z)/2$], the equation and the boundary
condition for $\psi$ is given by
\begin{equation}
  \label{eq:pde_psi}
  \p{\psi}{z} = b\p[2]{\psi}{x} \quad {\rm and} \quad
  \psi(0,z) = 0,  \quad  \psi\left(\pm\frac{L(z)}2,z\right) = \pm\frac{Q}{2}.
\end{equation}
We numerically integrate Eq.~\eref{eq:pde_psi} from $z=0$ using the
Crank-Nicholson method to obtain the prediction of the Kinematic
Model. 

Due to its continuum formulation, the Kinematic Model cannot predict
grain-level diffusion and mixing, so we now turn to statistical
kinematic models for the velocity profile, which postulate mechanisms
for random-packing dynamics.

\subsection{The Void Model}
\label{sub:void}

Since Eq.~\eref{eq:pde_v} has the form of a diffusion equation,
where the vertical distance $z$ plays the role of ``time'', it is
clear that any microscopic justification for the Kinematic Model
should be based on independent random walks. In fact, this is how the
model was first derived decades earlier, based on statistical
considerations. Although the continuum approach is more general, in
the sense that it is not tied to any specific microscopic mechanism,
it lacks a clear physical basis, so it is important to consider what
kind of microscopic mechanisms might support it.

Litwiniszyn first suggested the idea that particles are confined to a
fixed array of hypothetical ``cages" as they perform random walk from
one available cage to another during drainage~\cite{lit58, lit63a,
lit63b}. Then, Mullins~\cite{mullins72, mullins79} independently
proposed an equivalent model in terms of ``voids" rather than
particles, which is analogous to vacancy diffusion in crystals. In his
model, particles move passively downward in response to the passage of
voids, and the voids take directed random walks upward after emerging
from the orifice. 

Assuming that voids diffuse by non-interacting random walks, it is
straightforward to show that in continuum limit, at scales larger than
the grain diameter, the concentration (or probability density) of voids,
$\rho_v$, satisfies the diffusion equation,
\begin{equation}
  \label{eq:pde_rho}
  \p{\rho_v}{z} = b\p[2]{\rho_v}{x}.
\end{equation}
Since downward velocity $v$ is proportional to the frequency of the
void passage, this implies Eq.~\eref{eq:pde_v} of the Kinematic
Model. However, the equivalence of the two model assumes that voids
can be superimposed without interaction.

The void model also gives us an interpretation for the kinematic
parameter, $b$.  If a void undergoes a random horizontal displacement,
$\Delta x_v$, while it climbs up by $\Delta z_v$, the parameter $b$ is
given by 
\begin{equation}
  \label{eq:void_var}
  b = \frac{{\rm Var}(\Delta x_v)}{2\Delta z_v},
\end{equation}
which is the characteristic length of the void diffusion. However it
is very difficult to determine $b$ directly from
Eq.~\eref{eq:void_var}. $\Delta x_v$ and $\Delta z_v$ cannot be
measured from an experiment, nor does any \textit{a priori} choice
produce the measured value of $b$. Mullins also deduced $b\approx 2d$
from the velocity profile for round particles ($b\approx d/4$ for
irregular particles) without specifying the value of $\Delta x_v$ and
$\Delta z_v$. By contrast, Caram and Hong~\cite{caram91} assume a void
makes an one-to-one exchange with particles on a regular lattice when
they later revisited the void model. It is noteworthy that any regular
lattice of hard sphere packing under predicts $b$ ($b\ll d$) 
\cite{bazant04a}.

The void model faces more serious problems when it is used to predict
diffusion and mixing, which was not done by its proponents. If a
tracer particle is placed in a uniform flow driven by voids, the
particle does a directed random walk downward with precisely the same
diffusion length as the voids moving up. Thus particles are easily
mixed before they drop by a few particle diameters, which goes against
our everyday experience and experiments (see below).

\subsection{The Spot Model}
\label{sub:spot}
To address these contradictions, Bazant
\etal~\cite{bazant04a,bazant04b} proposed the Spot Model, which starts
from a mechanism for cooperative diffusion in a dense random
packing. It has roughly the same mean flow as in the kinematic model,
because it also assumes that particles move in response to upward
diffusing free volume, but this excess volume is carried in extended
``spots'' of slightly enhanced interstitial volume, not in voids.  

The kinematic parameter, $b$, is
now set by the diffusion length for spots,
\begin{equation}
  \label{eq:spot_var}
  b = \frac{{\rm Var}(\Delta x_s)}{2\Delta z_s},
\end{equation}
where, $\Delta x_s$ and $\Delta z_s$ are spot displacements in $x$ and
$z$ directions, respectively. Unlike a void which is a vacancy capable
of being filled by an entire particle, however, a spot carries small
fraction of interstitial space spread across an extended region and
causes all affected particle to move (on average) as a block with the
same displacement in the opposite direction to the spot. 

Of course, there are more complicated internal rearrangements, which
can be taken into account to achieve accurate spot-based
simulations~\cite{rycroft04}, but the simplest mathematical model
already captures many essential features of dense
drainage~\cite{bazant04a,bazant04b}. For example, it is easy to see
that the spot mechanism greatly reduces the diffusion length of
particles, compared to the diffusion length of free volume. Suppose
that a spot carries a total free volume $V_s$, and causes equal
displacements $(\Delta x_p,\; \Delta z_p)$, among $N_p$ particles of
volume $V_p$. The particle displacement can be related to the spot
displacement $(\Delta x_p,\; \Delta z_p)$ by an approximate expression
of total volume conservation,
\begin{equation}
  \label{eq:xp_xv}
  N_s\,V_p(\Delta x_p,\; \Delta z_p) = -V_s(\Delta x_s,\; \Delta z_s),
\end{equation}
which ignores boundary effects at the edge of the spot.
From this relation, we can  compute the particle diffusion length,
\begin{equation}
  \label{eq:bp_bv}
  b_p = \frac{{\rm Var}(\Delta x_p)}{2\Delta z_p}
  = \frac{w^2{\rm Var}(\Delta x_s)}{2w\Delta z_s} = w b_s
\end{equation}
which is smaller than the spot diffusion length by a factor, 
$w = V_s/N_pV_p$.  This  can in turn be related to the change, $\Delta
\phi$,
in local volume fraction, $\phi$, caused by the presence of the spot,
\begin{equation}
w = \frac{b_p}{b} = \frac{V_s}{N_p V_p} \approx \frac{\Delta
\phi^2}{\phi}
\end{equation}
It is well known from simulations and experiments that the volume
fraction fluctuates on the order of 1\% in a dense flow, so the Spot
Model thus predicts $w= b_p/b = O(10^{-2})$. (In our experiments, the
local area fraction of glass beads near the viewing wall varies by
less than three percent.)  The estimate of $w$ is further reduced by
noting that spots occur in large numbers and overlap, so that each
spot contributes only a small part of the change in local volume
fraction. We will test this prediction in our experiments.

\section{Experimental procedure}
\label{sec:exp}

\subsection{Experimental setup}
\label{sub:setup}

Our experimental apparatus and procedure is similar to that used in
our previous report~\cite{choi04}. We use black glass beads ($d=3.0\pm
0.1\,$mm) in a quasi-two-dimensional silo with length $L = 20.0\,$cm
($67d$), height $H = 90.0\,$cm ($300d$), and thickness $D = 2.5\,$cm
($8.3d$). The particles near the front wall of the silo are measured
through the transparent glass. The slight polydispersity reduces the
tendency for hexagonal packing to occur near the wall. The thickness
of the silo $D$ is large enough that finite-size effects are not
significant. We obtain similar results for both mean velocity and
diffusion when we increase $D$~\cite{choi04}. A distributed filling
procedure was used to fill the silo with the grains. The orifice is
opened and steady state flow is allowed to develop before acquiring
the images used for determining particle positions.

We view a rectangular region of $20.0 \times 50.0\,$cm above the
orifice with a resolution of $256 \times 1280$ pixels. Therefore each
particle diameter corresponds to $d=7.7$ pixels.  The images are
acquired at a rate of 125 frames per second. The camera memory allows
2048 consecutive images to be stored at this resolution and therefore
the maximum interval over which we can track a particle is about
16.4~s.

For the funnels in the hopper, plexi-glass wedges are placed on top of
the bottom plate. The surface property of wedge boundaries is
identical to the side walls. We use wedges with three different
angles, $\theta = 30^\circ, 45^\circ$ and $60^\circ$. The orifice size
$W = 18\,$mm is fixed for the hopper experiments while, it is varied
with $W = 12, 16$ and $20\,$mm for the silo. To gain good statistics,
three experiments are conducted for each funnel angle and orifice
size. We also use data from a wider range of orifice sizes than
acquired during our previous study~\cite{choi04} in
\sref{sub:flowrate}.

\subsection{Particle tracking}
\label{sub:tracking}
\begin{figure}
  \centering
  \parbox[t]{0.45\linewidth}{(a)\\
    \includegraphics[width=\linewidth]{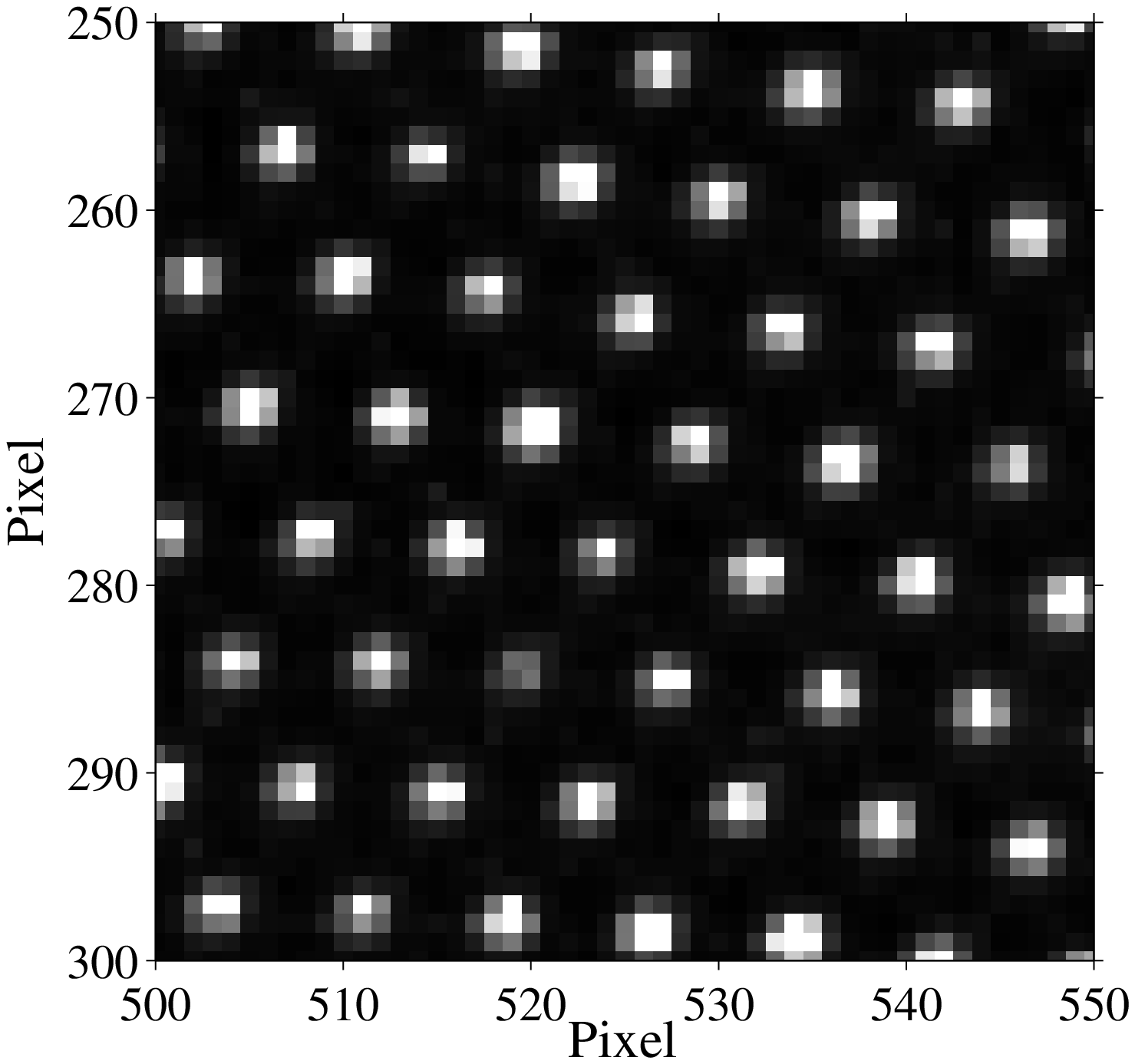}} \quad
  \parbox[t]{0.45\linewidth}{(b)\\
    \includegraphics[width=\linewidth]{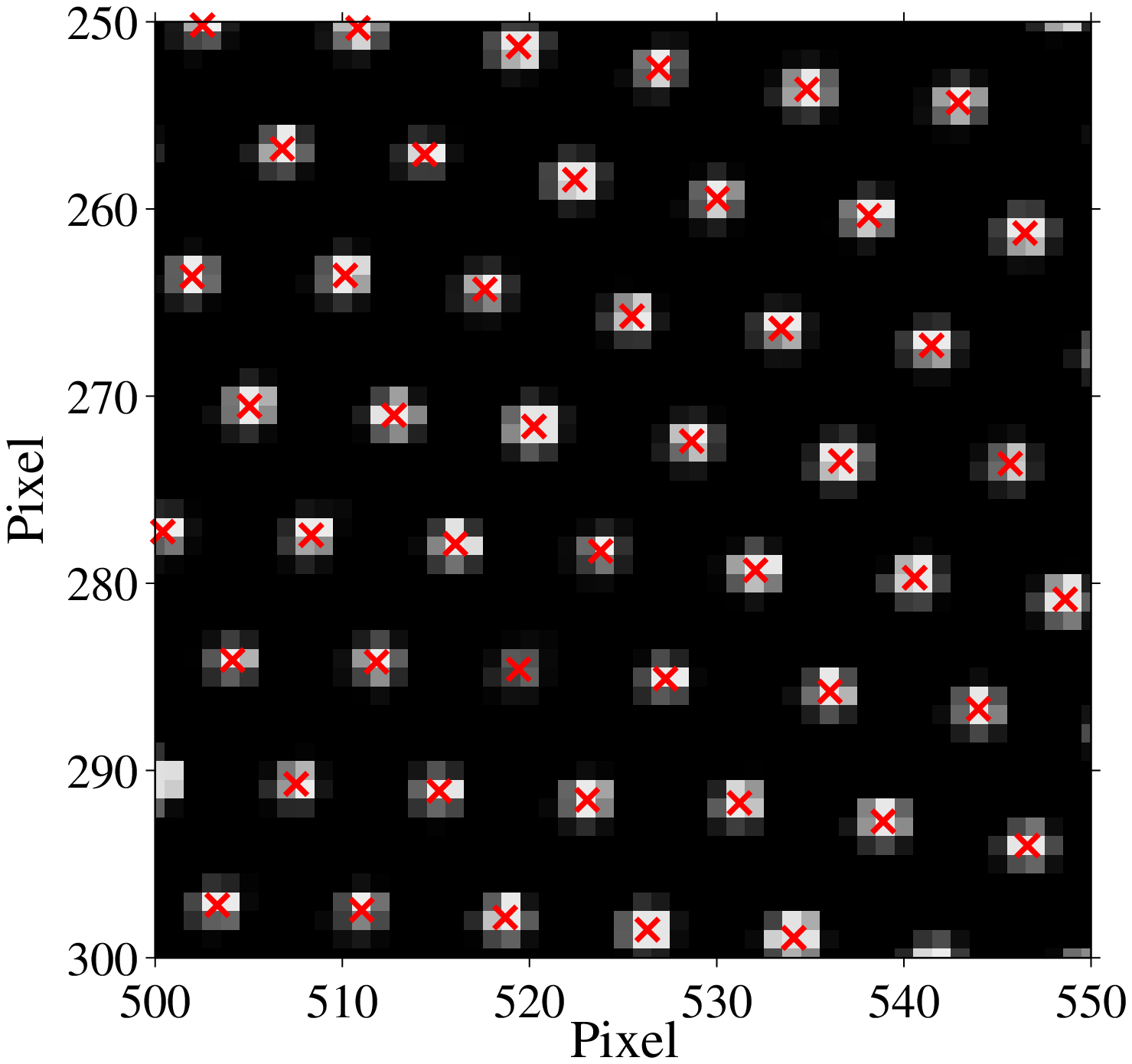}}
  \caption{\label{fig:track} (color online).  
    (a) A raw image of the glass beads acquired with 
    the high-speed camera, and (b) the preprocessed image along with the 
    position of the centroid of the identified particle ($\times$).}
\end{figure}

To identify the locations of particles from images, we employ the
algorithm proposed by Crocker and Weeks~\cite{crocker96}. In this
algorithm, the raw images are preprocessed to reduce the noise and the
background.  This involves convolving the image with a Gaussian filter
and then an average filter of roughly $d$ pixels respectively.  The
particle location is then identified with the centroid around the
local maximum brightness pixel in the modified image.  To optimize the
particle tracking for our experiment, the algorithm was also further
customized. Because the glass beads are circular, we use circular
shaped filter. We also set an intensity cutoff to discard the blur
images of particles located deep away from the front wall.  A sample
of an image before and after the processing is shown in
\fref{fig:track}. The position of the located particles is also superposed. 

After particles are located frame by frame, their trajectories should
be retrieved by ``connecting'' their positions in time. We associate a
particle in a frame with another in the next frame which is within a
radius of $0.66\,d$ pixels around the original position.  This simple
method works well avoiding more complicated multiple associations
except very near the orifice where the particles move more than
$0.66\,d$ pixels per frame.  The particles can be tracked there by
using a faster frame rate, but we do not do so here since bulk flow,
and not orifice dynamics, is the focus of our study.

\section{Analysis of the experimental results}
\label{sec:analysis}
\subsection{Comparison of the measured velocity profiles with the Kinematic Model}
\label{sub:meanvel}
\begin{figure}
  \centering
  \parbox[t]{0.34\linewidth}{(a)\\
    \includegraphics[width=\linewidth]{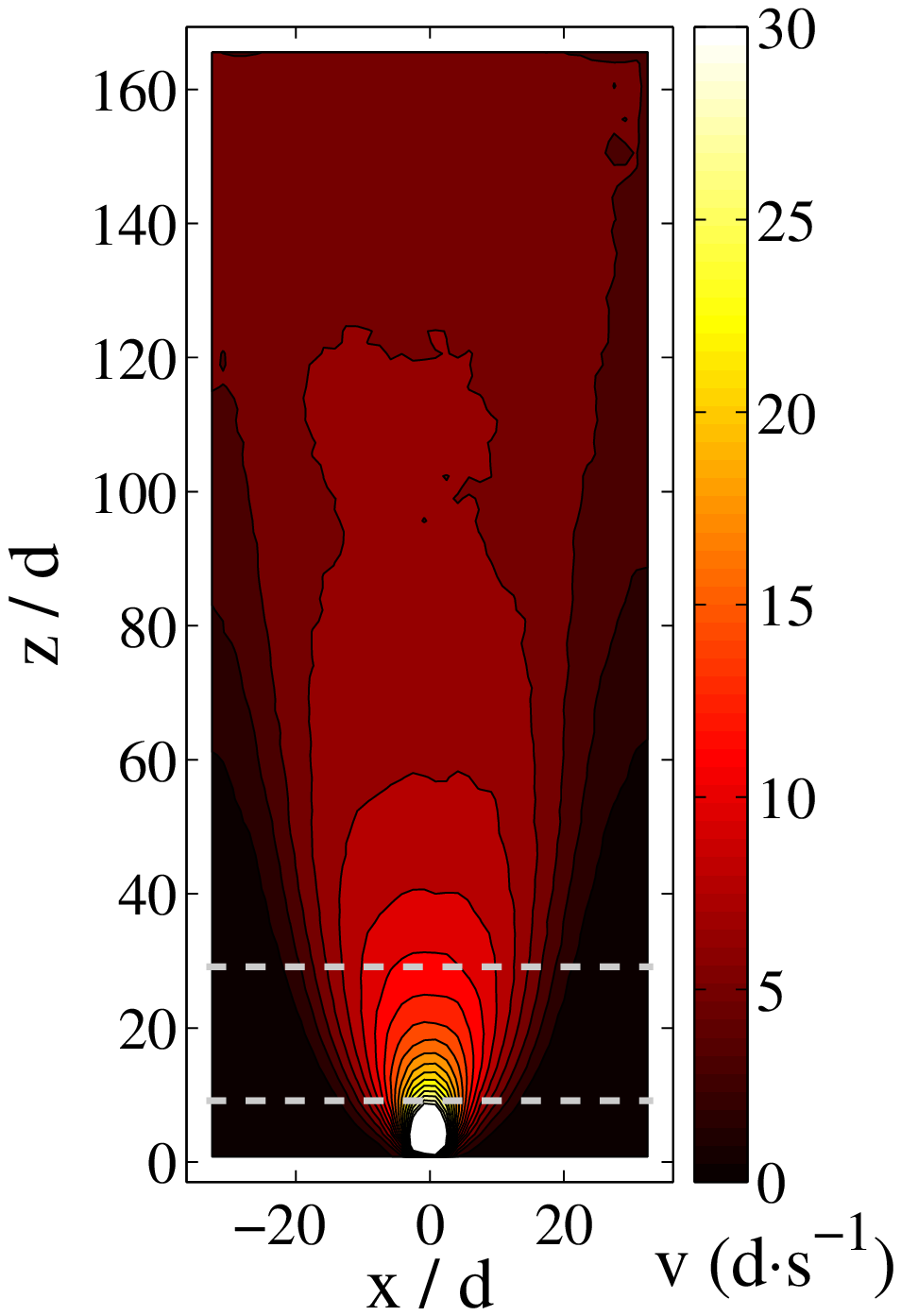}}
  \parbox[t]{0.63\linewidth}{(b)\\
  \includegraphics[width=\linewidth]{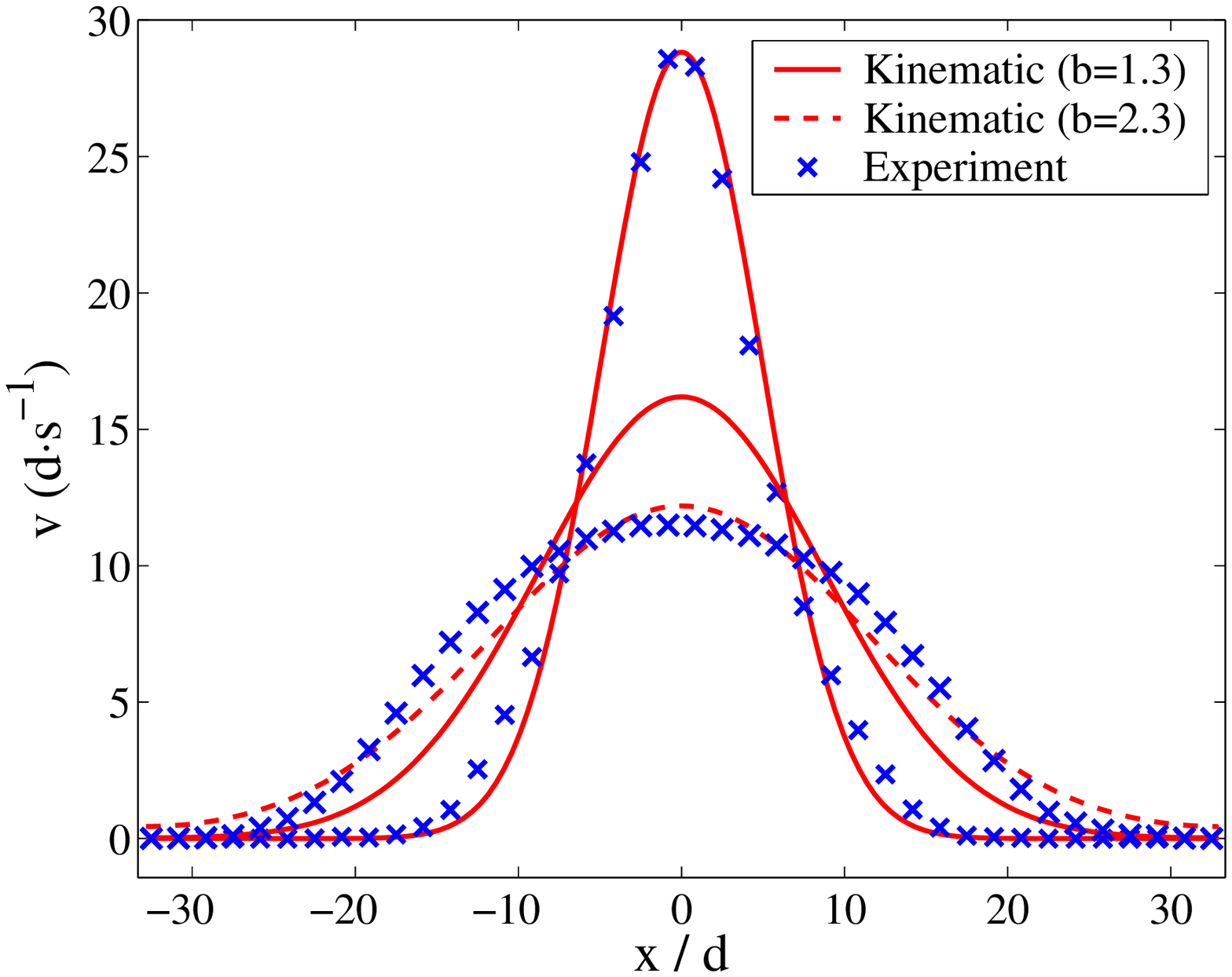}}
  \caption{\label{fig:deg00} (color online). 
    (a) Contour plot of the average downward velocity field, $v$ in 
    a flat-bottomed silo with an orifice width, $W = 16\,$mm. 
    (b) $v$ as a function of $x$ at the two heights, $z_1 = 9.1d$ and 
    $z_2 = 29.1d$ 
    indicated with gray dotted lines in (a). The result from the 
    Kinematic Model in the same geometry fits best with $b=1.3d$ for
    the $z_1$ profile, and $b=2.3d$ for the $z_2$ profile. 
    The result from the model for the $z_2$ profile with $b$ fitted at $z_1$  
    (narrow solid curve) is also shown.
  }
\end{figure}

\begin{figure}
  \centering
  \parbox[t]{0.34\linewidth}{(a)\\
    \includegraphics[width=\linewidth]{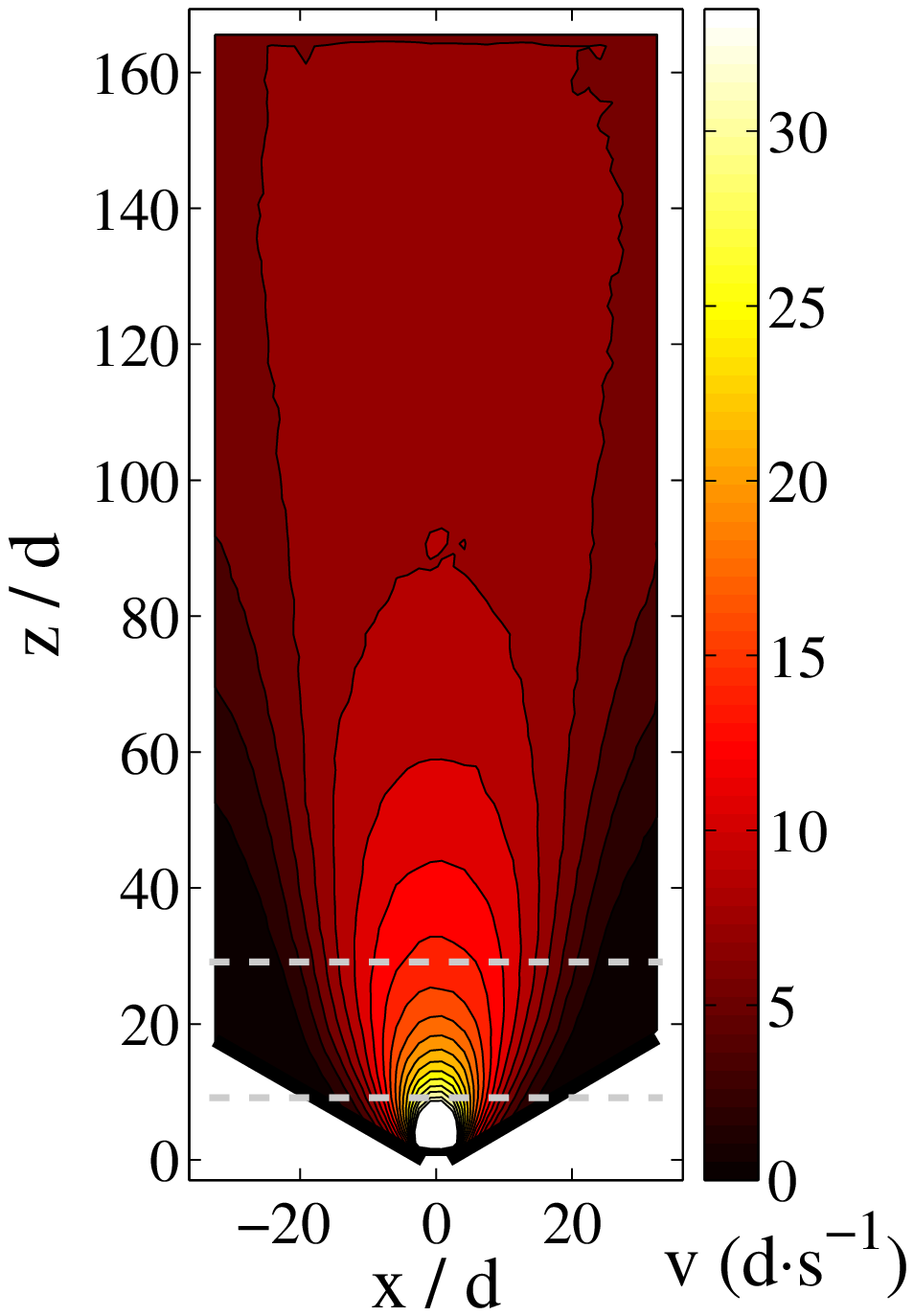}}
  \parbox[t]{0.63\linewidth}{(b)\\
  \includegraphics[width=\linewidth]{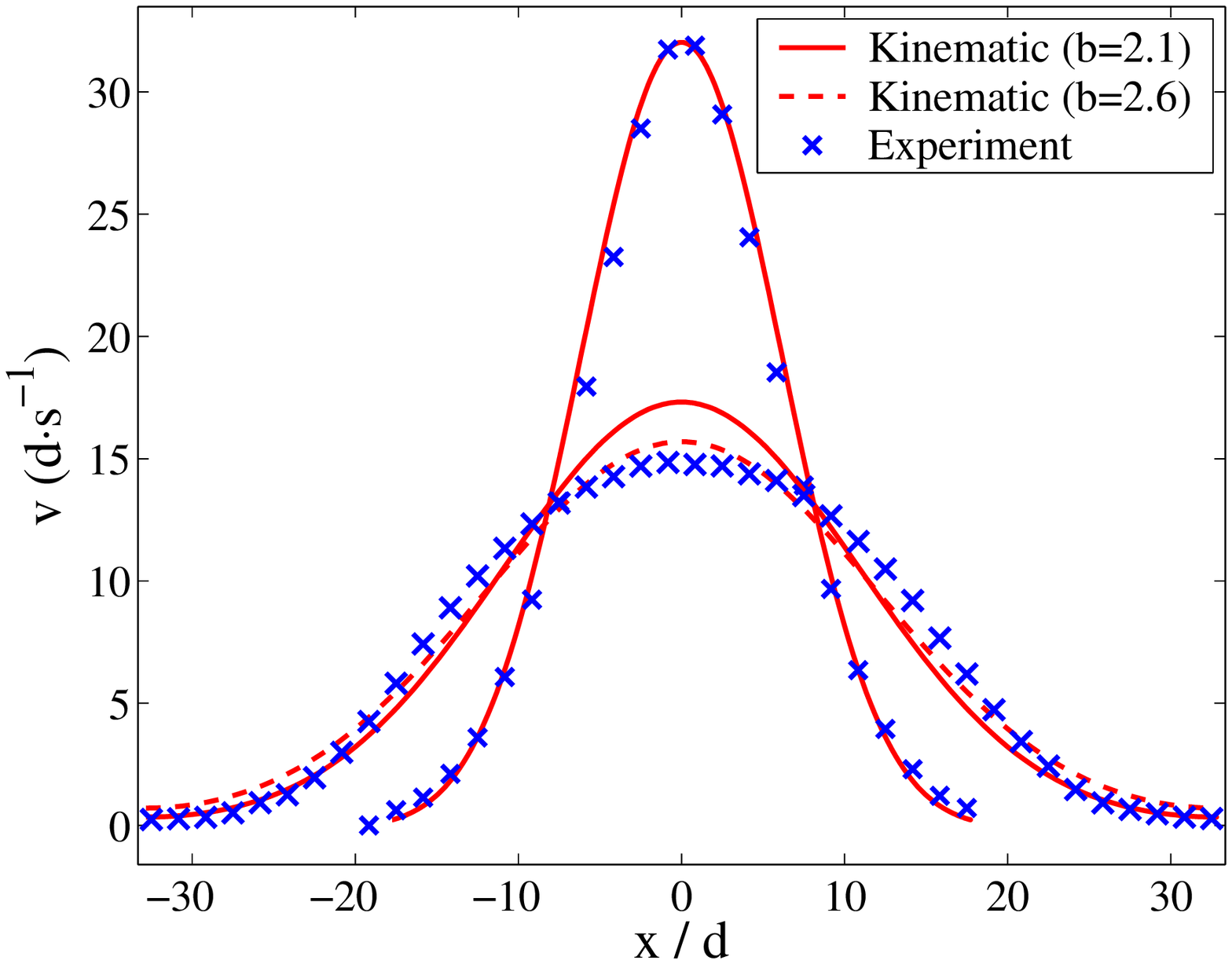}}
  \caption{\label{fig:deg30} (color online). 
    (a) Contour plot of the average downward velocity field, $v$ in 
    a hopper with angle, $\theta = 30^\circ$, and $W = 18\,$mm. 
    (b) $v$ as a function of $x$ at the two heights, $z_1 = 9.1d$ and
    $z_2 = 29.1d$ 
    indicated with gray dotted lines in (a). The result from the 
    Kinematic Model fits best with $b=2.1d$ for the $z_1$ profile 
    and $b=2.6d$ for the $z_2$ profile.
    The result from the model for the $z_2$ profile with $b$ fitted at $z_1$  
    (narrow solid curve) is also shown.
  }
\end{figure}

\begin{figure}
  \centering
  \parbox[t]{0.34\linewidth}{(a)\\
    \includegraphics[width=\linewidth]{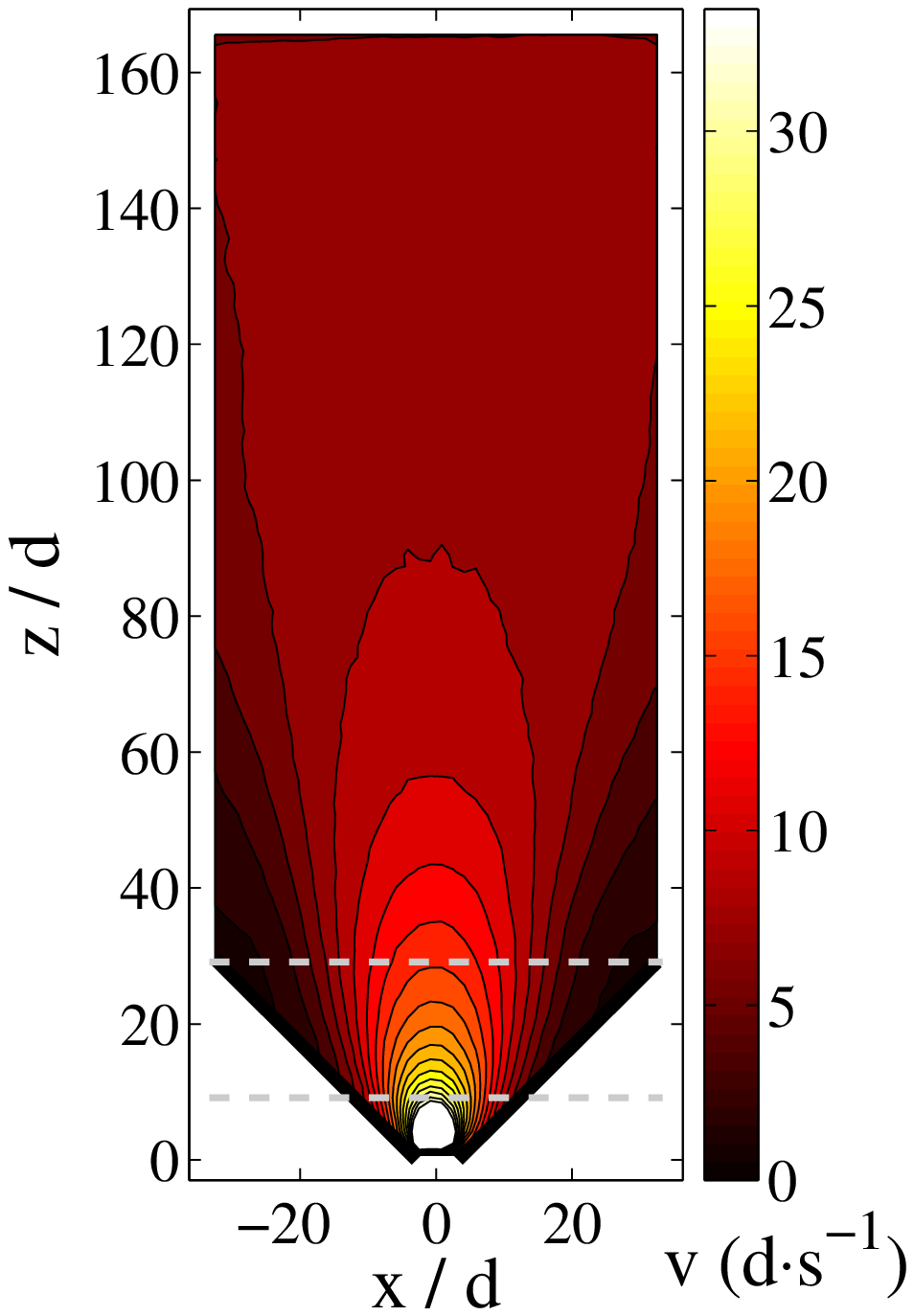}}
  \parbox[t]{0.63\linewidth}{(b)\\
  \includegraphics[width=\linewidth]{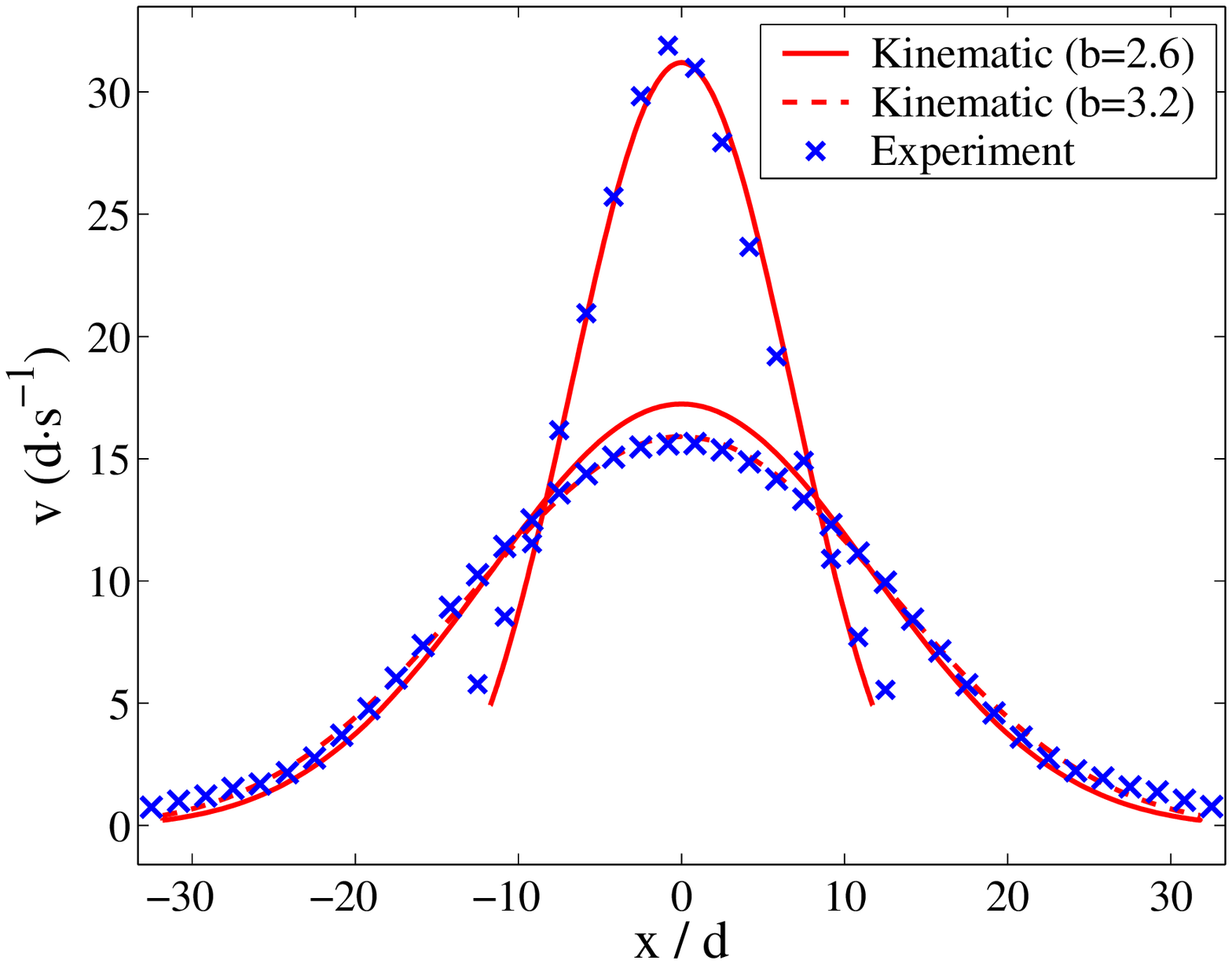}}
  \caption{\label{fig:deg45} (color online). 
    (a) Contour plot of the average downward velocity field, $v$ in 
    a hopper with angle, $\theta = 45^\circ$, and $W=18\,$mm. 
    (b) $v$ as a function of $x$ at the two heights, $z_1 = 9.1d$ and
    $z_2 = 29.1d$ 
    indicated with gray dotted lines in (a). The result from the 
    Kinematic Model fits best with $b=2.1d$ for the $z_1$ profile 
    and $b=2.6d$ for the $z_2$ profile.
    The result from the model for the $z_2$ profile with $b$ fitted at $z_1$  
    (narrow solid curve) is also shown.
  }
\end{figure}

\begin{figure}
  \centering
  \parbox[t]{0.34\linewidth}{(a)\\
    \includegraphics[width=\linewidth]{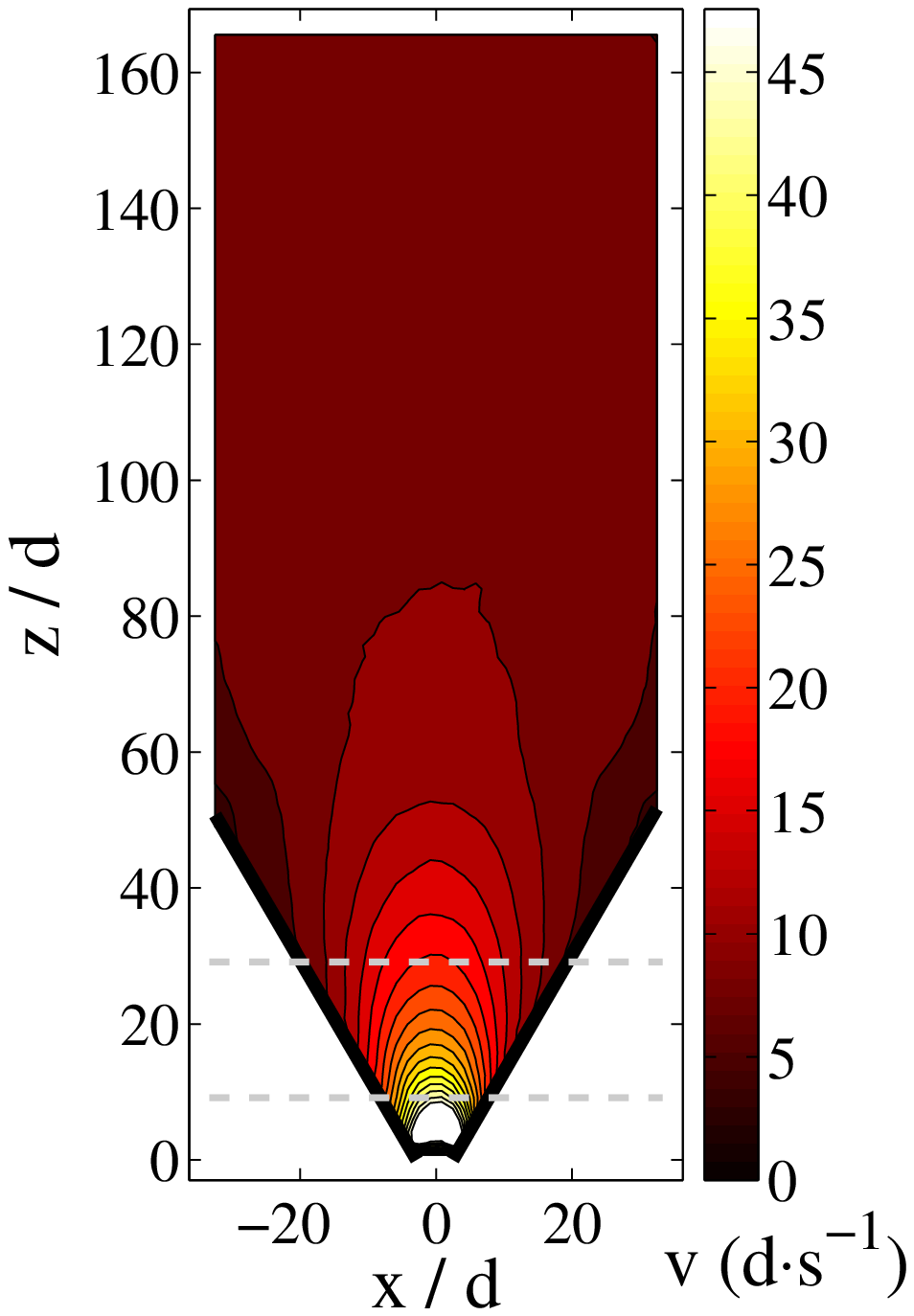}}
  \parbox[t]{0.63\linewidth}{(b)\\
  \includegraphics[width=\linewidth]{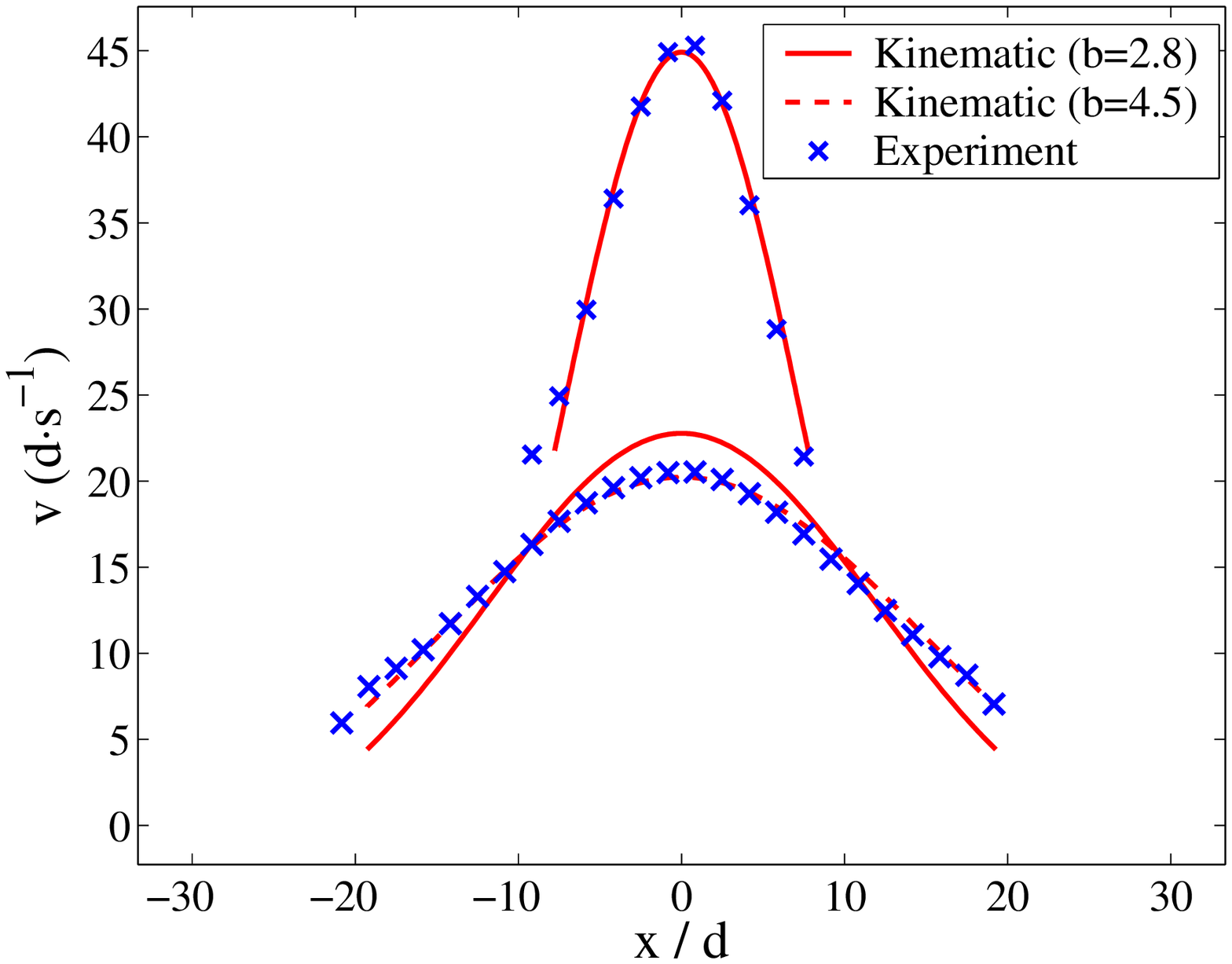}}
  \caption{\label{fig:deg60} (color online). 
    (a) Contour plot of the average downward velocity field, $v$ in 
    a hopper with angle, $\theta = 60^\circ$, and $W=18\,$mm. 
    (b) $v$ as a function of $x$ at the two heights, $z_1 = 9.1d$ and
    $z_2 = 29.1d$ 
    indicated as gray dotted lines in (a). The result from the 
    Kinematic Model fits best with $b=2.6d$ for the $z_1$ profile 
    and $b=3.2d$ for the $z_2$ profile.
    The result from the model for the $z_2$ profile with $b$ fitted at $z_1$  
    (narrow solid curve) is also shown.
  }
\end{figure}

We first compare the data from the flat-bottom silo with the model. 
\Fref{fig:deg00}(a) shows the contour plot of the average downward
velocity $v$. The mean velocity is obtained by dividing the observation 
window into square cells of size $1.6d \times 1.6d$. Then in each cell, 
the average is performed over the displacements of all the particles 
passing through the cell. We again take the average
of the field from three experiments. 
The data across experiments shows little variation,
which confirms that the velocity field is well-defined and stationary.
Thus we do not show the error bar in the plots of this paper unless
the concerned quantity has visible fluctuations.

The contour plot shows that $v$ is maximum right at the orifice and
appears to ``diffuse'' upward, in qualitative agreement with  the models
discussed above.  The regions in the left and right corner made by the
side walls and the bottom plate remain stagnant, and the boundary of
mobile region has a parabolic shape.  In \fref{fig:deg00}(b), we show
the profiles $v(x)$ at two cross sections $z_1 = 9.1d$ and $z_2 =
29.1d$ [dotted lines in \fref{fig:deg00}(a)] with the fit to the
Kinematic Model.  The diffusion length, $b=1.3d$ was the best fit for
the profile at $z_1$.  However, $b$ becomes larger when $z$ increases
as some previous reports have also
reported~\cite{nedderman92,medina98b}. The profile at $z_2$ is best
fit with $b=2.3d$, but it has a flattened shape at center with thinner
tail indicating further obvious deviations from the model.

The velocity profiles from the experiments with different orifice width
turn out to coincide when they are normalized by the flowrate as is
commonly observed in other dense granular flows~\cite{midi04}.
Thus the best fitting value of $b$ is independent of the flowrate. 
The dependence of the flowrate on the orifice width willl be discussed
in the next subsection.
 
We performed similar analysis of the experiments with the hoppers.
The contour plots along with the profiles at $z = z_1,\, z_2$ for the
angles, $30^\circ, 45^\circ$ and $60^\circ$ are presented in
\fref{fig:deg30},
\ref{fig:deg45} and \ref{fig:deg60} respectively.
As angle is increased, the stagnant region is diminished as the
particles slip on the wedge.  At $z_1$, the critical angle over which
slip occurs is between $30^\circ$ and $45^\circ$ and at $z_2$, it is
between $45^\circ$ and $60^\circ$. However the shape of equi-velocity
contours well above the funnel is not affected significantly by the
funnel's detailed shape.

The value of $b$ to obtain the best fit depends on the angle of the
hopper as well.  It increases from $2.1d$ to $2.8d$ for $z_1$, and
from $2.6d$ to $4.5d$ for $z_2$ as the angle is increased.

Although we observe some quantitative discrepancies with the simple Kinematic
Model with a constant coefficient, $b$, the flow is at least
qualitatively consistent. This appears not to be the case with
continuum models from critical-state soil
mechanics~\cite{nedderman92}, which generally predict sharp,
shock-like discontinuities in velocity (and stress, which we do not
measure) within the silo, especially near corners. We see no such
abrupt jumps in velocity in the silo,  only rather smooth velocity profiles.

\subsection{Flow rate dependence on the orifice size and the funnel angle}
\label{sub:flowrate}

\begin{figure}
  \centering
  \parbox[t]{0.55\linewidth}{(a)\\
    \includegraphics[width=\linewidth]{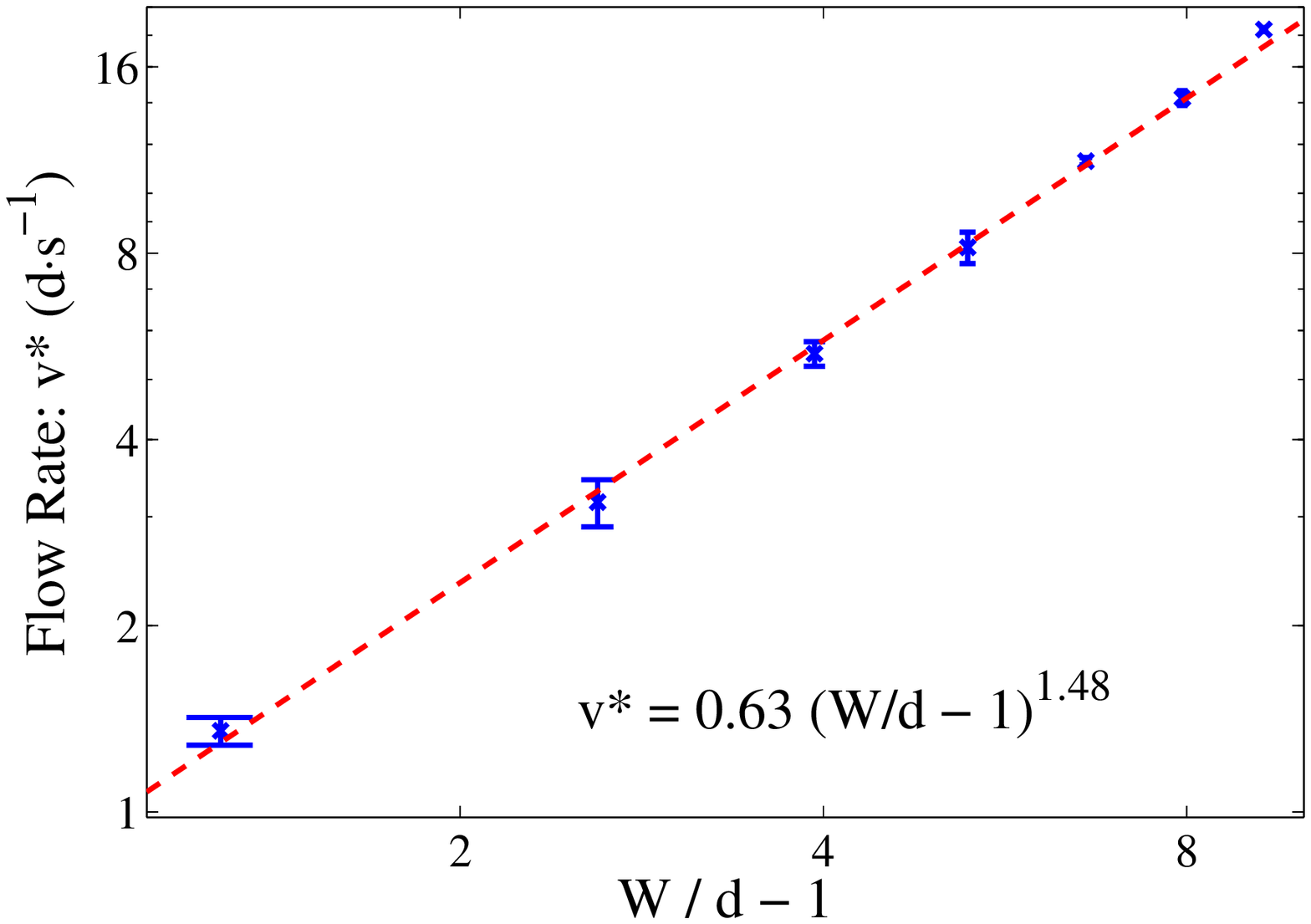}} \quad
  \parbox[t]{0.35\linewidth}{(b)\\
    \includegraphics[width=\linewidth]{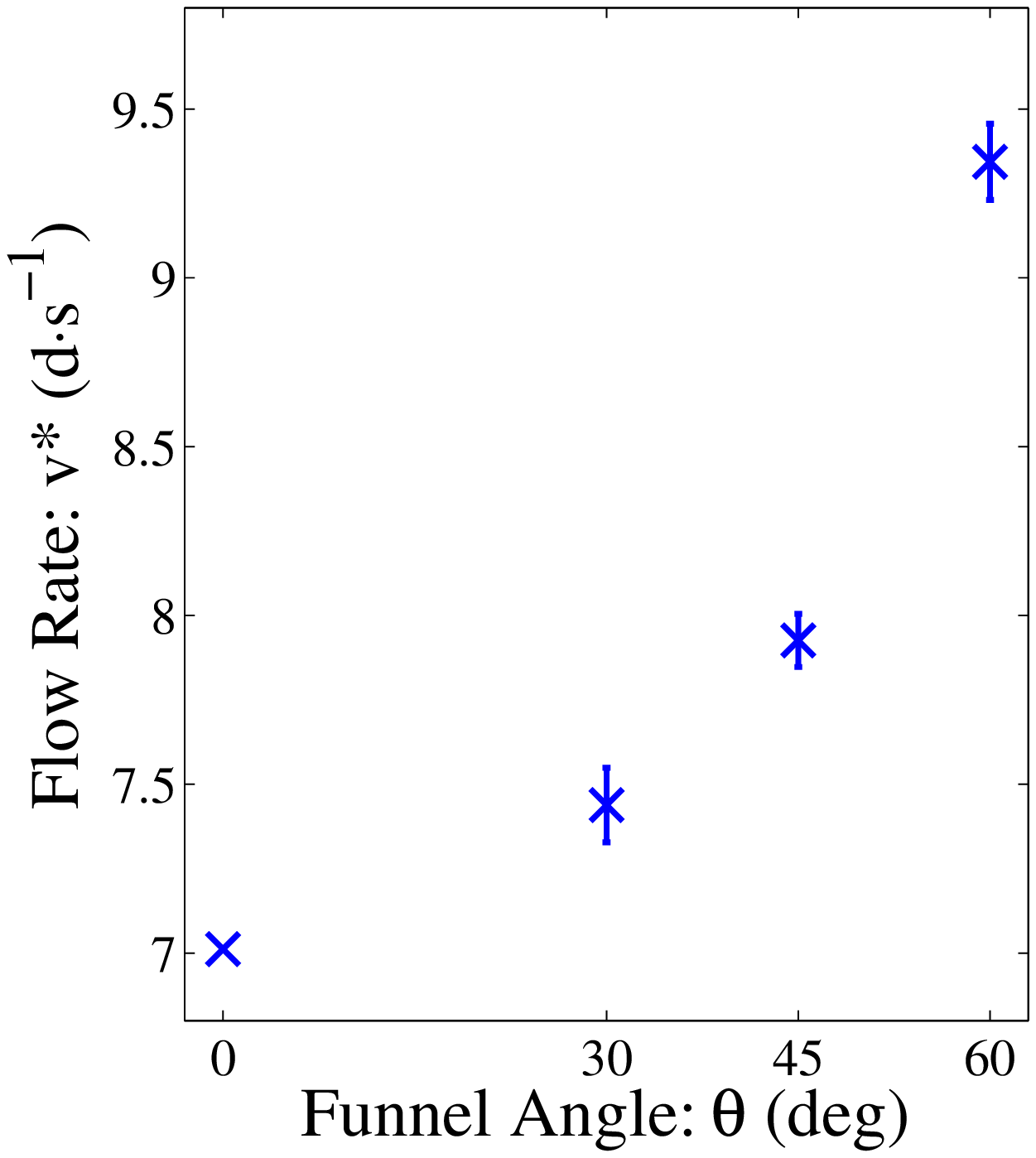}}
  \caption{\label{fig:flowrate} (color online). 
    The dependence of flow rate on (a) the effective orifice width,
    $W/d-1$ in a flat-bottomed silo (log-log scale), and 
    (b) the funnel angle $\theta$ in a hopper with a fixed orifice width.
    The flow rate is measured averaging the downward velocity in the
    plug-flow region. The fitting of (a) validates the result of 
    a dimensional analysis, $Q \propto (W-d)^{1.5}$.
  }
\end{figure}

The mass flow rate in a silo during discharge was an important
subject of early research. Using drainage experiments in
cylindrical silos with a circular orifice, Beverloo,
\etal~\cite{beverloo61} reported a relation known as the Beverloo
correlation
\begin{equation}
  Q \propto \rho\sqrt{g}\;(W - kd)^{2.5},\quad k = 1.4
\end{equation}
where $Q$ is the mass flow rate, $\rho$ is the bulk density of
packing, $g$ is the gravitational constant and $W$ is the diameter of
orifice.  It is usually argued that $Q \propto
\rho\sqrt{g}\;(W-d)^{2.5}$ is the only form which can be deduced from
the dimensional analysis as $(W-d)$ is the effective diameter (or
width) where particle centers can be placed within the orifice, but
arching and other effects could also introduce the particle diameter
$d$ and thus another dimensionless parameter, $d/W$. Instead, the
Beverloo correlation includes a somewhat controversial factor $W-kd$,
where the empirical factor $k$ is claimed to derive from the region
near the orifice rim which obstructs the passage of particles.  This
picture could be consistent with the concept of an ``empty annulus"
proposed by Brown and Richards~\cite{brown70}.

For a \textit{slit} orifice with a quasi-two-dimensional silo as in
our experiment, the dependence can be obtained to be
\begin{equation}
  Q \propto \rho\sqrt{g}\;(D-d)\,(W-kd)^{1.5},
\end{equation}
because the flow rate is linear with system depth $D-d$
\footnote{It should be noted that the orifice in our system is entirely open
from front to back surface. Thus the ``empty annulus" argument is
difficult to apply in the direction of silo depth. We cannot find the
exact dependence on $D$ because we fix $D=2.5\,$cm.}. We investigated
the flow rate dependence on orifice width using our data. Although the
discharged mass flux is not directly measured, we use the overall
average velocity, $v^* = Q/L$ to obtain the flow rate.

\Fref{fig:flowrate}(a) shows the relation between the flow rate and
orifice size in log-log scale. When $k=1$, the data fits 
to a power law scaling with an exponent of $1.48$.
Although $k=0.94$ gives the exact exponent of $1.5$, we do not attach
much importance to the deviation as our flow rate measure is indirect.
However, it is sufficient to check that the Beverloo correlation (dimensional
analysis) holds in a 2-D silo. 

We also investigated how the funnel angle affects the flow rate.
In order to compare the rate at a fixed orifice width, 
we interpolate the rate with $W = 18\,$mm  from data with
$W=16,\,20$ and $24\,$mm for the silo experiment. 
\Fref{fig:flowrate}(b) shows a consistent increase in the flow rate 
as the angle increases. The flow rate in the $60^\circ$ funnel turns 
out to be about 33\% more than that in the flat-bottom silo. 
This dependence is consistent with the data from Ref.~\cite{tuzun82a},
although the reported increase of the flow rate is smaller than our data.
We believe the increased flow rate largely comes from the fact that
the smooth rigid boundary facilitates the passage of particles. 
As seen clearly from \fref{fig:deg00}(b)$\sim$\ref{fig:deg60}(b),
the stagnant zone present in the corners of flat-bottom silo gets
replaced by wedges. Thus the slip velocity at the boundary
increases as the angles increases, which make the out-going flow at 
the orifice ($z=0$) more uniform and shear-free. This effect appears 
to allow the particles to exit the orifice more easily.  

\subsection{Diffusion of particles in an uniform flow}
\label{sub:diffusion}
As explained in \sref{sec:model}, particle diffusion is a key property
to distinguish between different possible microscopic mechanisms for
dense granular flow. The Void Model and the Spot Model predict quite
similar mean flow profiles (given by the Kinematic Model scales much
larger than the grain size), but the former predicts $b_p/b = O(1)$
while the latter predicts $b_p/b = O(10^{-2})$. In this section, we
briefly discuss measurements of particle diffusion in our experiments,
as also previously reported in Ref.~\cite{choi04}.

To measure diffusion, the random component of
particle displacement is obtained by subtracting the average component:
\begin{equation}
 \dx = \dx' - u \dt \quad {\rm and} \quad  \dz = \dz' + v \dt,
\end{equation}
where $\dx'$ (or $\dz'$) is the observed displacement in
$x$ (or $z$) direction, $\dx$ (or $\dz$) is the random displacement
in the same direction, and $\dt$ is the time gap between two consecutive
frames and can be increased by any integer multiple. The observation window is 
in a nearly plug-flow region far from the orifice, where $u$ is negligible 
and $v$ is almost uniform (and set by varying the orifice width).

The probability density distributions of $\dx$ and $\dz$ 
are observed to display fat tails compared to a Gaussian distribution. The statistics of 
$\dz$ also show an anisotropy due to gravitational acceleration
and inelastic collisions. When the width of the distributions is 
examined as a function of $\dt$, the scaling shows
a crossover from super-diffusion, $\dxsq
\propto \dt^{1.5}\/$ and $\dzsq \propto \dt^{1.6}\/$, to diffusion,
$\dxsq \propto \dzsq \propto \dt$. A significant observation is that
the lines of $\dxsq$ and $\dzsq$ for different $v$ collapse into a
single line when they are plotted against the distance dropped,
$v\dt$, allowing us to characterize the dynamics only by distance
moved, independent of the flow rate, $v$. We found that this dynamical
crossover occurs after a particle falls roughly by its diameter
irrespective of the flow rate.  

The fact that the dynamics only depends on geometry strongly suggests
that advection and diffusion have the same physical source (such as a
the passage of a void or spot). It also suggests that structural
rearrangements with long-lasting contacts dominate diffusion in dense
granular flows, as opposed to ballistic collisions, which are central
to the kinetic theory of gases. A direct evidence is that the
cage-breaking length is estimated to be order $100d$ from the rate of
the nearest neighbor loss~\cite{choi04}. Our results suggest that the
concept of ``granular temperature'' based on thermodynamic,
randomizing collisions is of dubious value in slow, dense granular
flows.

Since the free volume models correctly predict the geometry dominated
diffusion, we can proceed to evaluate them quantitatively. We compute
the P\'eclet number, the dimensionless ratio of advection to
diffusion, defined as
\begin{equation}
  \Pe_x = \lim_{\dt\to\infty}\frac{2Vd\dt}{\dxsq} = \frac{d}{b_{p,x}}
  \quad {\rm and} \quad
  \Pe_z = \lim_{\dt\to\infty}\frac{2Vd\dt}{\dzsq} = \frac{d}{b_{p,z}}
\end{equation}
are interpreted as the distances (in unit of $d$) for a particle to
fall before it diffuses by a diameter in $x$ or $z$ direction,
respectively. The large measured values, $\Pe_x = 320$ and $\Pe_z =
150$, indicate that advection dominates diffusion. Since $b/d \approx
2$, we also find $w_x = b_{p,x}/b \approx 1/600$ and $b_{p,z}/b
\approx 1/300$, which is consistent with the simple
prediction of the Spot Model, $w \approx 10^{-3} - 10^{-2}$.  Of
course, the data firmly rejects the Void Model, which predicts, $w
\approx 1$, and cage breaking at the scale of one particle diameter. 

\section{Discussion}
\label{sec:discussion}

\begin{figure}
  \centering
  \parbox[t]{0.7\linewidth}{
    \includegraphics[width=\linewidth]{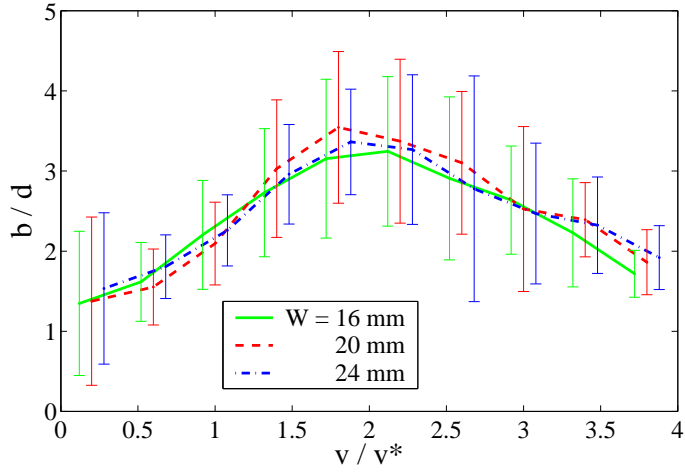}}
  \caption{\label{fig:b_v} (color online). 
    The locally measured diffusion length, $b$
    as a function of the normalized velocity, $v/v^*$.
  }
\end{figure}

In \sref{sub:meanvel}, we observed that the Kinematic Model with a constant
parameter $b$ is not consistent with the  experiments. It was found that $b$ depends
on the height and the funnel angle. In this section, we investigate the
validity of two important assumptions of the Kinematic Model,
namely the constitutive law \eref{eq:consti}, and 
the generalized boundary condition \eref{eq:bdd} 
for the funnel geometry.

First, we directly check the constitutive law \eref{eq:consti}
using the results from our experiments. In each cell that was used for 
averaging the velocity, we measure the horizontal velocity $u$, 
the downward velocity gradient $\partial v / \partial x$, and 
therefore the diffusion length $b$.
\Fref{fig:b_v} shows the distribution of the locally measured values of
$b$. As expected, it shows a wide fluctuation scattered from $b=d$ to
$b=3d$. When $b$ is associated with $v$, we find $b$ increases upto
$3.4d$ and decrease as $v$ increases. In other words, we observe
higher $b$ moving away from the stagnant zone and towards the fast
flow regions at the center. However, for the fastest-flow regions 
close to the orifice, $b$ decreases.
A reasonable implication of the increase in $b$ is that the slightly 
lower density in the fast-flow region due to dilation makes horizontal 
movements easier. The decrease in $b$ at higher $v$ is perhaps related to 
the fact that particles undergo collisional flow in the fast flowing 
regions near the orifice. 
Since the particles are less locked to neighbors than in the dense bulk
away from the orifice, the shear in the downward velocity result in  
less horizontal movement, therefore, smaller $b$.

A few further comments about \fref{fig:b_v} are in order. To collect
meaningful statistics for $b$, we ignore shear-free zones
(e.g. stagnant zone and plug-flow regions where the gradient of $v$ is
negligible), where $b$ is likely to have large errors.  We accomplish
this by only considering cells where gradient is larger than $5\%$ of
the characteristic magnitude, $v^*/d$, where $v^*$ is overall average
velocity in the plug region. Although we only discuss $b$ for the silo
experiments in \fref{fig:b_v}, a similar trend is also found for the
hoppers as well.

The correlation between $v$ and $b$ gives some clues to explain the
discrepancies in \sref{sub:meanvel}. The Kinematic Model with constant
$b$ fails to capture the development of a more plug-like plateau in
the velocity profile even with larger values of $b$. However, if
higher $b$ is applied to the region around the center (where $v$ is
high), and lower $b$ is applied to the region close to walls (where
$v$ is low), the model would come into closer agreement with
experiment. 

In an effort to understand the universality of this pattern, we use
the overall average velocity, $v^*$, to normalize the downward
velocity, $v$, from different flow rates (or orifice size). As shown in
\fref{fig:b_v}, pairs of $(v/v^*,b)$ for three different flow rates
fall into nearly the same pattern. This is a consistent with the
trends observed in Ref.~\cite{choi04} that increasing the flow rate
merely \textit{fast-forwards} the entire dynamics, without
changing the geometrical sequence of events.

Our way to describe our experimental results {\it a posteriori} is via
a modified constitutive law with a variable diffusion length, $b$,
which depends on the (scaled) local velocity:
\begin{equation}
  \label{eq:nonlinear} 
 u = b \p{v}{x} \quad {\rm and} \quad  b = b^* \Phi\left(\frac{v}{v^*}\right),
\end{equation}
where $b^*$ is an effective diffusion length and
$\Phi$ is a dimensionless scaling function. Note that the velocity
field satisfying Eq.~\eref{eq:nonlinear} is still linear with respect
to rescaling the total magnitude of the velocity (by changing the
total flow rate) since $v/v^*$ is invariant when $v$ is
rescaled. However, the velocity profile in space is governed by  a
nonlinear diffusion equation,
\begin{equation}
\p{v}{z} = b^* \p{
}{x}\left[\Phi\left(\frac{v}{v^*}\right)\p{v}{x}\right]
\end{equation}
It is well known that spreading solutions to this equation (analogous
to a concentration-dependent diffusivity) are flatter in the central
region (compared to a Gaussian) when $\Phi$ is an increasing function
of its argument ~\cite{crank}.

We should consider what might be the microscopic reason for a
nonlinear diffusion length in the Kinematic Model. In general, it
would arise from interactions between different spots, which are
neglected as a first approximation. It makes sense that spots of free
volume should diffuse less when they find themselves in a more slowly
flowing, less dense, region, with fewer other nearby spots. This could
explain why $b$ appears to grow with velocity (or spot
concentration). On the other hand, the flow in the upper part of the
silo becomes more plug-like should exhibit less diffusion than the
lower region of greater shear near the orifice, so it remains unclear
whether the nonlinear model \eref{eq:nonlinear} can be given a firm
microscopic justification. Further comparison with theory and
experiment is needed to settle this question.

\begin{figure}
  \centering \parbox[t]{0.7\linewidth}{
  \includegraphics[width=\linewidth]{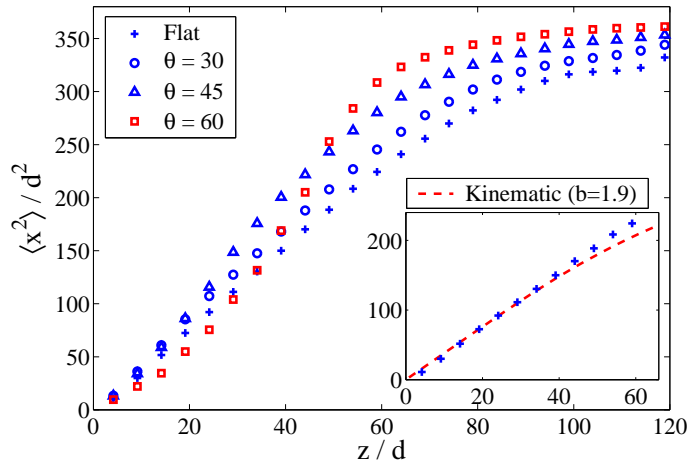}}
  \caption{\label{fig:var_z} (color online).  The variance (squared
  effective width) of the downward velocity profile, $\langle
  x^2\rangle_v$, as a function of the vertical coordinate, $z$ for
 different funnel angles.  }
\label{fig:var}
\end{figure}

The next issue to test is the boundary condition at the side
walls. Specifically, it is important to test if the model can be
simply extended from open silos to hoppers by using Eq.~\eref{eq:bdd}.
It is interesting to note that the curvature of the profile at $z=z_1$
around $x=0$ remains the same for the different funnel angles [see
\fref{fig:deg00}(b)$\sim$\ref{fig:deg60}(b)].  In fact, it is $b$
that should increase from $b = 1.3d$ to $b = 2.8d$ in order to
reproduce the same curvature as the hopper angle is increased.  For a
more quantitative argument, we show in \fref{fig:var_z} the variance
of the downward velocity profile  (a measure of its squared width),  
\begin{equation}
\langle  x^2\rangle_v
= \frac{\int x^2\, v(x)dx }{\int v(x)dx}
\end{equation} 
as a function of height, $z$.  From Eq.~\eref{eq:spot_var}, the slope
of the linear regime near the orifice is equal to $2b$, and the value
of the implied $b$ does not significantly vary from $b=1.9d$ for the
silo, as can be seen in the inset to \fref{fig:var}.

We conclude, therefore, that extending the Kinematic Model to a hopper
with non-vertical walls does not seem to be successful with the naive
idea of Eq.~\eref{eq:bdd}, which assumes the same bulk constitutive
law holds all the way to the boundary. 

It may be that a nonlinear constitutive law as in
Eq.~\eref{eq:nonlinear} can improve the situation because particles
slip more on a funnel wall and $b$ thus tends to be higher than in the
silo. However, there may still be problems higher in the tank where
the flowing region meets the vertical side walls. The boundary
condition \eref{eq:bdd} requires that the strain rate (horizontal
gradient of vertical velocity) vanishes at a vertical wall, and yet
small velocity gradients are observed near the walls in the upper
region in Figs. 2-5. We plan to compare the nonlinear Kinematic Model,
as well as other continuum models from critical-state  mechanics
and hourglass theory, more closely with the experimental flow profiles
in future work.

\section{Summary}
\label{sec:summary}

In summary, we have used high speed imaging techniques to track the
positions of granular materials draining inside silos and hoppers. We
compared our data with the continuum Kinetic Model and two possible
microscopic theories which predict similar mean flow, the Void Model
and Spot Model. These models are appealing due to their mathematical
simplicity and completeness, which allows direct application to
various geometries. The models also predict smooth velocity profiles,
free of shock-like discontinuities, quite consistent with the
experiments. Systematic deviations are observed, implying various
assumptions, such as a constant diffusion length, are too simple to
capture all aspects of the flow profile, but it may be that
modifications can be made to improve the agreement.  For example, we
infer that the kinematic parameter, $b$, increases with the local
velocity, which would imply that the spot diffusion length increases
in the presence of other spots. Still, it is clear that further work
is also needed to develop boundary conditions for both discrete
and continuous models of slow, dense granular flows.

\section*{Acknowledgments}
This work was supported by the U.S. Department of Energy (Grant
No. DE-FG02-02ER25530) and the Norbert Weiner Research Fund and the
NEC Fund at MIT, and the National Science Foundation (Grant
No. CTS-0334587) at Clark University.

\end{document}